\newcommand{\be}{\begin{equation}}
\newcommand{\ee}{\end{equation}}
\newcommand{\bea}{\begin{eqnarray}}
\newcommand{\eea}{\end{eqnarray}}

\newcommand{\nn}{\nonumber}

\def\p{\partial }

\def\a{\alpha }

\def\td{\tilde }
\def\s{\sigma }

\def\ep{\gb }

\def\ga{\alpha }
\def\gb{\beta }
\def\gam{\gamma }
\def\td{\tilde }

\def\I{{\rm I}}
\def\II{{\rm II}}
\def\ai{ \theta }

\documentclass[12pt]{article}

\input epsf
\usepackage{amsmath,amsfonts,epsfig,color,latexsym}
\usepackage{amssymb}
\usepackage{amsfonts}
\usepackage{amssymb}
\renewcommand{\thefootnote}{\fnsymbol{footnote}}

\def\appendix#1{
  \addtocounter{section}{1}
  \setcounter{equation}{0}
  \renewcommand{\thesection}{\Alph{section}}
  \section*{Appendix \thesection\protect\indent \parbox[t]{11.15cm}
  {#1} }
  \addcontentsline{toc}{section}{Appendix \thesection\ \ \ #1}
  }

\def \td {\tilde}

\begin{document}


\null\vskip-24pt 
\hfill {\tt hep-th/0601072}
\vskip0.2truecm
\begin{center}
\vskip 0.2truecm {\Large\bf
Handbook on string  decay }
\vskip 0.2truecm

\vskip 0.7truecm
\vskip 0.7truecm

{\bf Roberto Iengo$^a$ and Jorge G. Russo$^b$}\\
\vskip 0.4truecm
\vskip 0.4truecm

${}^a${\it  International School for Advanced Studies (SISSA)\\
Via Beirut 2-4, I-34013 Trieste, Italy} \\
{\it  INFN, Sezione di Trieste}

\medskip

$^{b}${\it 
Instituci\' o Catalana de Recerca i Estudis Avan\c{c}ats (ICREA),\\
Departament ECM,
Facultat de F\'\i sica, Universitat de Barcelona,
 Spain} 
\end{center}
\vskip 0.2truecm 

\noindent\centerline{\bf Abstract}

We explain simple semi-classical rules to estimate the lifetime
of any given  highly-excited quantum state of the  string spectrum in
flat spacetime.
We discuss both the decays by splitting into two massive states and by massless emission.
As an application,
we study a solution describing a rotating and pulsating
ellipse which becomes folded
at an instant of time -- the ``squashing ellipse''.
This string interpolates between the folded string with maximum angular momentum and the
pulsating circular string. 
We explicitly compute the quantum decay rate for the corresponding
quantum state, and verify the basic rules that we propose.
Finally, we give a more general
(4-parameter) family of closed string
solutions representing rotating and pulsating elliptical strings.

\newpage

\renewcommand{\thefootnote}{\arabic{footnote}}
\setcounter{footnote}{0}


\section{Decay rate due to breaking}
\setcounter{equation}{0}

In general, a string can decay either by emitting light particles or
by  splitting into two massive strings. 
Which channel is dominant depends on the string state
and on the number of uncompact dimensions.
In the situation of splitting into two massive strings with masses
much greater
than $l_s^{-1}$, the two outgoing strings are highly excited string
states which admit a classical description.
The decay rate in this case can be estimated by semiclassical arguments.
In this section we will discuss the different situations in detail and give 
an estimate of the decay rate due to breaking in each case.

\subsection{Open strings}

{} The probability of breaking for an open string
was studied in \cite{Dai}. There it was suggested that it is
constant along the string and proportional to $g_o^2$.
We propose that, more precisely,
the probability per unit time of breaking on a given point and at a
given instant  is
\be
P_o={g_o^2 \over T} \ ,
\label{buno}
\ee
where $T$ is the period of oscillations of the string
(so the probability of breaking in one period is $TP_o=g_o^2$
in agreement with \cite{Dai}).

In the  gauge $X_0= {T\over \pi}\, \tau $, we have
$$
M={1\over 2\pi\a' }\int_0^{\pi }d\sigma \, \dot X_0={T\over 2\pi\a' }\
,
$$
or $T=2\pi \a' \, M$. This formula applies to both open and closed strings.
The decay rate for an open string is thus obtained by multiplying
$P_o$ by the number of points of the string $L/l_s$, where
$l_s=\sqrt{\a '} $ and $L$ is the length of the string, and by the number
of ``instants'' in one period $T/l_s$. We find
\be
\Gamma_{\rm open}\cong  \Big( {L\over l_s} \Big)\, \Big( {T\over l_s}
\Big)\ P_o \cong g_o^2 \, {L \over l_s^2}\ ,
\label{bdos}
\ee
in agreement with \cite{Dai}. This formula can be checked against the
explicit
quantum calculation of the decay rate 
for the open string
with maximum angular momentum performed in \cite{Okada}. 
It was found that $\Gamma_{\rm open}\cong g_o^2  M$, where
 $M=L/l_s^2$ for this particular string state,
in precise agreement with (\ref{bdos}). 
We have reproduced this quantum calculation independently.
Figure 1a shows 
$\Gamma_{\rm open}(M)$ obtained from the exact evaluation of ${\rm
Im}(\Delta M^2)$ at one loop.

\subsection{Closed strings}

In closed string theory,
in the absence of D-branes, a closed string can only break into two
outgoing closed strings. This means that the breaking is possible only
if two points of the string meet.
The possible configurations are:

\begin{itemize}

\item The closed string is folded, with two sides in permanent
contact.

\item The closed string becomes folded only at some instant of time.

\item The closed string has two points which are in permanent contact.

\item The closed string has two points which get in contact only at 
some instant of time.

\item The distance between any two points of the string is $\gg l_s $
at all times. In this case the string is ``unbreakable'', meaning that
the decay rate into two massive ($M_{1,2}\gg l_s^{-1}$) string states
is exponentially suppressed, $\Gamma = O(e^{-cM^2})$. 

\end{itemize}

We will assume that the probability of breaking at a given point once
in a period is
given by $P_o={g_o^2 \over T} $, as proposed above. 
This is a natural assumption since
the breaking is a local phenomenon which should not depend
on global properties of the string, i.e. whether the string is closed
or open. In addition, for the breaking to take place, it is necessary that
the string  simultaneously breaks at
two points which are in contact. Here `simultaneously' means
within an interval of order $l_s$.

In what follows we apply these simple rules and estimate the decay
rate for all the cases listed above.

\subsubsection{Folded string}

The two folds of the closed string are in permanent contact,
so the string can break at any time.
We can estimate the decay rate by viewing it as
two open strings on the top of each other.
The breaking can take place only if at a given time the 
two open strings break at the same point, up to an uncertainty of order
$l_s$.

The decay rate is thus
\be
\Gamma_{\rm folded}\cong  \Big( {L\over l_s} \Big)\, \Big( {T\over l_s}
\Big)\ P_o \, P_o l_s \cong g_s^2 \, {L \over l_s T }
\cong g_s^2 \, {L \over  l_s^3 M}
\ ,\qquad g_s=g_o^2\ .
\label{jhgs}
\ee
In particular, for the folded string with maximum angular momentum ${\rm J_{max}}$,
one has $L\cong l_s^2 M$, so that 
\be
\Gamma_{\rm J_{max}}
\cong g_s^2 \,   l_s^{-1}\ .
\label{jhg}
\ee
Remarkably, it is constant independent of $M$. This estimate applies
at large $N=\a' M^2/4$.
This can be compared with the exact quantum calculation of the decay rate.
This was computed in \cite{CIR}, and here we have added more numerical
data, up to $N=119$. Figure 1b shows the decay rate as a function of
the mass. We see that it asymptotically approaches to a constant,
confirming the semiclassical estimate.

\bigskip

\begin{figure}[ht]
\centerline{
\epsfig{file=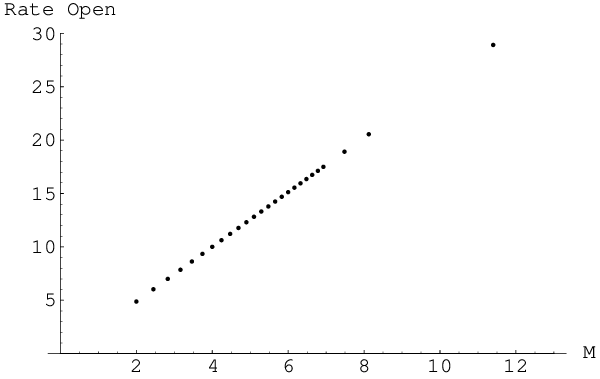,width=.5\textwidth}
1a
\qquad
\epsfig{file=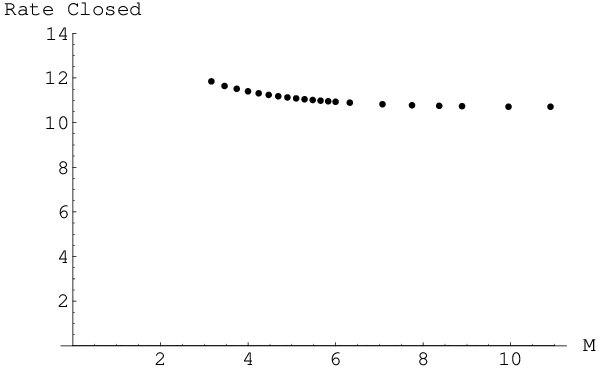,width=.5\textwidth}
1b }
 \caption{\it Decay rate versus mass of the open and closed strings with maximum angular
 momentum.
Figure 1a: Open string. Figure 1b: Closed string. }
\end{figure}

\bigskip

 The decay rates of figure 1a and 1b contain also the channels  where
 one of the emitted particles is massless. 
 The semiclassical estimate, however, assumes that both the final
 string fragments are massive.
%
%
Therefore, it is of interest to compare the classical prediction with the 
numerical results restricted to the channels where both decay products
are massive.  

For the case of the open string the behavior of the rate is quite the
same as for the total rate.
For the case of the closed string there is something more to comment.

In figs.2a and 2b we report the decay rate 
restricted to the massive channels for
the closed string with maximal angular momentum with two possible 
assumptions for the target spacetime.
Fig.2a is the case in which every dimension is uncompact. Fig.2b is the
case in which six space dimensions are toroidally compactified,
the string lying in an uncompact plane (the decay products are taken
to have zero winding and KK modes in the compact space).

We see that the decay rate due to massive emission shown in fig.2a is
significantly lower than the total decay rate shown in fig.1b.
This shows that the
decay rate due to massless emission is important (and, at least in
this range, decreasing).
Now let us compare to the case of $D=4$.
We see that flattening towards a constant of the behavior of the massive-channels-lifetime is more evident in Fig.2b  
(the compactified case) rather than in Fig.2a (with nine uncompact
spatial dimensions).
It is likely that this is a kinematical effect due to the
phase-space being much wider in $D=10$  than in $D=4$
uncompact dimensions. As a result, in $D=10$, getting to the asymptotic regime
requires larger masses to
overcome the opposite tendency of the phase-space opening.  
We have also checked that the massless contribution to the decay rate in $D=4$ is larger
by a factor $\sim 20$ than the massive contribution, and it is
increasing with the mass. This is the expected behavior in $D=4$, as explained in section 2.

\begin{figure}[ht]
\centerline{
\epsfig{file=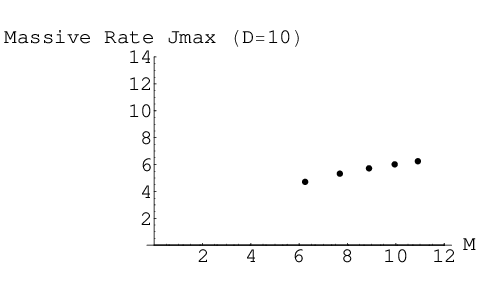,width=.5\textwidth}
2a
\qquad
\epsfig{file=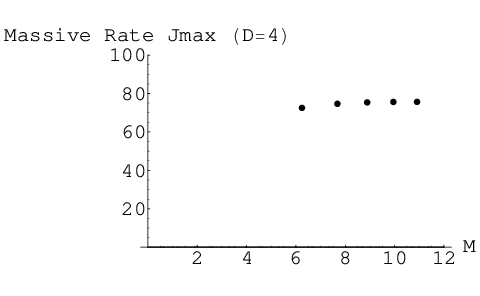,width=.5\textwidth}
2b }
 \caption{\it Decay rate versus mass of the closed strings with
 maximum angular momentum, including massive channels only.
Figure 2a: D=10. Figure 2b: D=4.}
\end{figure}

Another interesting case is that of a folded string which
is wound $n$ times. In this case, $L=l_s \sqrt{N}/n$.
The decay rate (\ref{jhgs}) must be multiplied by a factor ${1\over
2}\ 2n(2n-1)$,
which is the number of possible pairings between $2n$ folds of the
string. This gives
\be
\Gamma_{\rm 2n-folded}
\cong g_s^2 \, l_s^{-1} (2n-1) \ .
\label{jhgss}
\ee

\subsubsection{String which becomes folded at an instant of time}

This is the case of the pulsating and rotating 
ellipse of section 3, the squashing ellipse. 
We are interested in the decay rate due to breaking of the string,
i.e. into two massive strings. This process is only possible at the
instant
when the ellipse is completely squashed so that different points of the closed
strings
get in contact. The time interval where the breaking is possible is of
order $l_s$ (see section 3).
By definition, the decay rate is the number of events at each period
of oscillation of the string. This means that the decay rate will be
given by
\be
\Gamma_{\rm squash} \cong {l_s\over T} \ \Gamma_{\rm folded} \cong
{1\over\sqrt{N} }\ \Gamma_{\rm folded} \cong
 g_s^2 \, {L \over  l_s^4M^2}\ .
\ee
In particular, for the specific squashing string of section 3, one
has $L\sim l_s^2 M$, so that
\be
\Gamma_{\rm squash} \cong
 g_s^2 \, {1 \over l_s \sqrt{N} }\ .
\label{sq}
\ee
We will find the same law by explicit calculation of
the quantum decay in section 3. This is a non-trivial confirmation of 
the semiclassical rules given above.

In Fig.3 we report the results of that computation, in the form of the
lifetime (that is, the inverse of the rate) versus mass, restricted to the decay channels  
where both fragments are massive, following the discussion made for 
the case of the maximal angular momentum. According to (\ref{sq}), one
expects a linear behavior, lifetime $\cong {\rm const.}\ M$.
{}From fig.3, we see that the quantum calculation reproduces the
classical linear behavior
even for  rather small masses both for $D=10$ and
$D=4$ uncompact spacetime dimensions.

\medskip

\begin{figure}[ht]
\centerline{
\epsfig{file=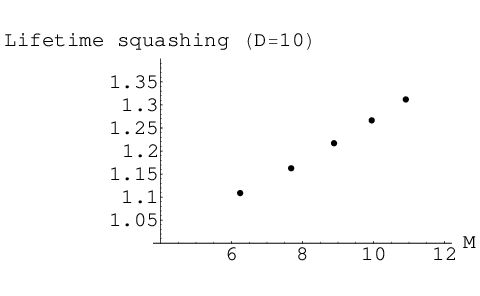,width=.5\textwidth}
3a
\qquad
\epsfig{file=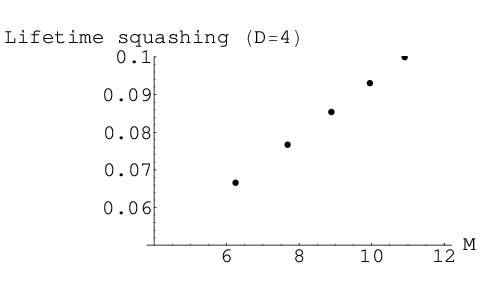,width=.5\textwidth}
3b }
 \caption{\it Lifetime versus mass of the squashing strings,including  massive channels only.
Figure 3a: D=10. Figure 3b: D=4.}
\end{figure}

\medskip

\subsubsection{Pulsating circular string}

This is a circular string which shrinks to a point once in a period.
The decay rate can be calculated just as in the squashing string case,
but now taking into account that the two points where the breaking
takes place are arbitrary, since by the time the string is completely shrunk all points 
are in contact. This means that there is an
additional
factor of $L/l_s$. Thus we get
\be
\Gamma_{\rm pulsating} \cong
 g_s^2 \, {L^2 \over M^2 l_s^5}\cong  g_s^2 \,   l_s^{-1} \ ,
\ee
where we have used that $L\cong l_s^2M$ for the pulsating circular string.
Thus the decay rate is the same as in the folded string case with
maximum angular momentum.
We have verified this remarkable fact 
by the exact quantum calculation (see section 3).

\subsubsection{Strings with two points in permanent contact}

Let us call this particular closed string configuration a 
``crossed" closed string.
This is similar to the folded string, but now we do not have to multiply
by the number of points of the string $L/l_s$ .
Thus we get
\be
\Gamma_{\rm crossed} 
\cong g_s^2 \, {1\over  l_s^2M} \cong
 g_s^2 \, {1 \over l_s \sqrt{N} }
\ ,
\label{mkl}
\ee
which is similar to the decay rate of the squashing ellipse (\ref{sq}).

\subsubsection{Strings with a finite number of contacts in one period}

Let us now consider a string where two points  get in contact 
at an instant of time. 
The decay rate is obtained by adding a factor $l_s/T$ to the previous
result (\ref{mkl}).
This factor represents the fraction of time that the strings are in contact,
which is the quantum spread $l_s$ over the period $T$. Thus
\be
\Gamma \cong  g_s^2 \ {l_s\over  T^2 } \cong g_s^2 \ {1\over  l_s\, N }\ .
\label{mkls}
\ee
{}For these strings the decay rate due to breaking (\ref{mkls})
is very small and they are expected to decay primarily by massless radiation.

This case also puts a bound on the decay rate due to splitting
of an average closed string state.
An average string state can be thought of as a random walk process
(see e.g. \cite{turok,manes}) and
the crossing of the string can occur at most a finite number of times in one period of oscillation.
To get the decay rate into massive string states  one should multiply the result (\ref{mkls})
by the probability that any two points meet in one period. 
Since this number is less than one, this means that, in general, 
$\bar \Gamma_{\rm massive}< g_s^2 {1\over l_sN }$.
For these strings, the decay due to massless radiation is largely
dominant (see below).

\section{Decay rate due to massless emission}
\setcounter{equation}{0}

The quantum massless emission from a closed string contains
contributions from four sectors:  NS-NS, R-NS, NS-R and R-R.
Computing the massless decay rate for a generic quantum string state
is in general complicated. In particular, the covariant vertex
operator for a general massive string state is not known.
We have explicitly computed the decay rate for every channel,
including massless emission, in the cases of  the string with
maximum angular momentum Jmax and for the rotating ring
\cite{CIR,CIR2}, and for the squashing string (section 3).

The rotating ring is a circular string rotating in
two orthogonal planes (see the corresponding classical solution in
section 4.1).
In this case, we found \cite{CIR2} that NS-NS emission (which includes graviton,
dilaton and antisymmetric tensor) is dominant, whereas  R-NS, NS-R 
emissions  are suppressed by a factor $1/N$ and R-R emission is suppressed
by a factor $1/N^ 2$, where $N=\a' M^ 2/4$ and $M$ is the mass of the
decaying string state.
Moreover, in this case the NS-NS emission can be accurately described
as a radiation process from a classical antenna represented by the
classical rotating ring solution.

The classical radiation from a source $T_{\mu\nu}(x_0,\vec x)$ in $D$
uncompact spacetime dimensions is given by
\be
\Gamma =g_s^ 2{\omega^{D-3}\over M^2} \int d^{D-2} \Omega\ \sum_{\xi ,\bar\xi }\ |J|^2\ ,
\label{gamrad}
\ee
$$
J=\int dx_0 d^{D-1} \vec x \ e^{i \omega  X_0 -i\vec p .\vec X}  \xi^\mu \tilde \xi ^\nu\ T_ {\mu\nu}(X_0,\vec X)\ .
$$
For a classical string solution, the energy momentum tensor is
\be
T_{\mu\nu} =\int d\sigma d\tau \delta^{(d)} (x-X(\tau,\sigma )) 
\p X_\mu \bar\p X_\nu \ .
\ee
so that $|J|^ 2=|J_R|^ 2|J_L|^ 2$ with (gauge $\xi^ 0=\vec\xi\cdot
\vec p=0$)
\be
J_R=\int_0^ {2\pi } d\sigma \ 
e^{i p_- X_{R+}(\s)}\  
\vec\xi\cdot  \partial \vec X_R^ T(\s)
\label{jayr}
\ee
\be
J_L=\int_0^ {2\pi } d\sigma \ 
e^{i p_- X_{L+}(\s )}\  
\vec\xi\cdot \partial \vec X_L^ T(\s)
\label{jayl}
\ee
where we have chosen the frame where the momentum of the emitted
massless particle is $p^ \mu =(\omega,-\omega,\vec 0)$,
and the gauge $X_0=\a' M\tau =2\sqrt{\a'N}\tau  $. $X_{\pm }$ refer to
light-cone coordinates where $p_+=0$.
The radiated energy is $\omega ={M^ 2-{M'}^ 2\over 2M}=N_0/(2\sqrt{N})$,
where we have set $\a'=4$ and $N_0\equiv N-N'$, with  ${M}=\sqrt{N}$
being  the mass of the original state
and ${M'}=\sqrt{N'}$ the mass of the  
massive state after the emission.

\medskip

It is important to understand when
the classical formula represents a good approximation in order to
compute the decay rate due to massless radiation. In other words,
which are the string states that radiate like a classical antenna.

The classical formula is expected to hold for $\omega \ll
O(1/\sqrt{\a'})$, i.e. for $N_0\ll \sqrt{N}$. 
In this regime we also expect that NS-R, R-NS and R-R emission is small.
If massless NS-NS emission with higher energies is suppressed, then the
classical formula can be used to compute the total radiation emission.

In general, since  $i p_- X_{R,L+}(\s  )\sim i N_0 f_{R,L}(\s )$ in the
exponent in (\ref{jayr}),  (\ref{jayl}),  
then $J_{R,L}$ are exponentially suppressed as a function of
$N_0$, unless there is a saddle point in the integration over $\s
$ in eqs.(\ref{jayr}), (\ref{jayl}), or a ``kink''   in the function
$f_{R,L}(\s )$ . 
A saddle point (``cusp'') occurs if  $\partial_\s X_{R+}$
and  $\partial_\s X_{L+}$  vanish for  some $\s$, while a kink occurs
when there is  discontinuity
in the first derivative in the function $f_{R,L}(\s  )$.
This is the case of the cusp and kink string configurations studied in
\cite{vilenkin,DV}. 

So let us first assume that there is no cusp or kink. 
In this case, 
$$
J_{R,L}= M\ h_{R,L}(N_0,\Omega )
$$
where $h_{R,L}(N_0,\Omega )$ are exponentially suppressed for $N_0\gg 1$.
We have used the fact that $\vec\xi\cdot \partial \vec X_{R,L}^ T$ is
generically proportional to $M$ as can be seen from the
Virasoro constraints.
Therefore, the decay rate (\ref{gamrad}) is given by
\be
\Gamma \cong {\rm const.}\ g_s^ 2 M^ {5-D} \, N_0^ {D-3}\int d^{D-2} \Omega\
|h_R (N_0,\Omega )h_L (N_0,\Omega )|^ 2
\ee
The total rate is obtained by summing over $N_0$ (i.e. all possible
energies of the massless particle). This sum is convergent and one
obtains
\be
\Gamma_{\rm total \ rad} \cong {\rm const.}\ g_s^ 2 M^ {5-D}
\ee
This law has been verified explicitly  in \cite{CIR2} by both the classical and quantum
calculation for the rotating ring, which has no cusp or kink.
In this case, since decays into massive channels are exponentially
suppressed, $\Gamma_{\rm total\ rad}\cong  \Gamma_{\rm total} $, so 
the lifetime of the ring state is  $\sim   g_s^ {-2}\ M^ {D-5}$.
Note that the time for the state to radiate away all of its energy is
much longer by a factor proportional to $N=M^ 2$.

\medskip

Let us now consider string configurations with cusps. Examples of
these are the string with maximum angular momentum $J_{\rm max}$, the
squashing string and the open string.
In the  $J_{\rm max}$ or the
squashing string cases, $ \partial \vec X_{R,L}^ T$ vanish for some $\s
$, which can be taken at $\s =0 $, behaving as $ \partial \vec X_{R,L}^ T\sim \s $. Using
the Virasoro constraint, $X_{R,L+}\sim\s^ 3$. It follows that \cite{vilenkin,DV}
\be
J_{R,L}^{\rm cusp}\sim N^ {1/2} \ N_0^ {\ -2/3}\ .
\label{csp}
\ee
Note that the condition of the existence of a cusp depends on the
definition of $X_+$, which depends on a particular  $p^ \mu
=(\omega,-\omega,\vec 0)$. Therefore the behavior (\ref{csp}) holds
for a particular angle.
Then the classical decay rate (\ref{gamrad}) on this angle becomes
\be
\Gamma (N_0)\sim {\rm const.}\ g_s^ 2 M^ {5-D}\ N_0^ {3D-17\over3}\ .
\ee
The sum over $N_0$ is convergent only for $D\leq 4$. Therefore, in higher
dimensions, in the presence of such cusps, the classical formula cannot be used.
In $D=4$, the classical formula for the spectrum is reliable only in
the range  $N_0\ll\sqrt{N}$ 
(note that the computation of the average value of the emitted energy, proportional to $\langle N_0\rangle $,
would be divergent even in $D=4$ at the particular angle for which (\ref{csp}) holds).
A similar analysis can be carried out for more general string
configutations with cusps  for which  $ \partial \vec X_{R,L}^ T\sim
\s^ \beta $. This includes the kink case when $\beta =0$. In this case 
, where $ \partial \vec X_{R,L}^ T$ is discontinuous,  one gets
\be
J_{R,L}^{\rm kink}\sim N^ {1/2} \ N_0^ { -1}\ ,\qquad \Gamma (N_0)\sim {\rm const.}\ g_s^ 2 M^ {5-D}\ N_0^ {D-7}\ .
\ee
String configurations with kinks appear naturally as a result of the splitting of a string \cite{IR2}.
The kink or cusp feature is preserved during the classical evolution of a string.
{}However, from the above formulas  we see that kink and cusp string configurations have a radiation spectrum 
which does not decay exponentially. As a result, such strings radiate more, providing a dissipation mechanism  
which should smooth out the kinks and cusps quantum mechanically.

\medskip

Whenever the  sum over $N_0$ is convergent for every angle one obtains again, either in the cusp or kink case, 
$\Gamma_{\rm total \ rad} \cong {\rm const.}\ g_s^ 2 M^ {5-D}$. By
comparing with the exact numerical results of the full quantum
computation, we
have verified that this is indeed the behavior in $D=4$ for Jmax and
the squashing string, while this is not the behavior of $\Gamma_{\rm total \ rad}$ in $D=10$, as expected.

\medskip

Finally, another case where the massless radiation emission can be
computed explicitly is that of an average string state.
In this case, one finds \cite{AR,CIR2}
\be
\bar \Gamma_{\rm total\ rad}\cong g_s^2 M
\ee
This formula is the same even if some dimensions are compact.
It was also derived in \cite{DV} from a study of the mass shift.

In the previous section we have made an estimate for  the decay rate due
to breaking for an average string state $\bar \Gamma_{\rm massive}<
g_s^2 {1\over l_sN }$.
This is much smaller than $\bar \Gamma_{\rm total\ rad}$ so 
for an average string state  the lifetime will be determined by the massless radiation
channel,
i.e.  given by $ {\cal T}=(g_s^2 M)^{-1}$.

\section{The squashing ellipse}
\setcounter{equation}{0}

The classical string solution is given by
\bea
x_1 &=& 2L \cos \ai \, \cos\tau\, \cos\s  \ ,\qquad x_2=2L\sin\tau\,
\cos\s   \ ,
\nonumber\\
 x_3 &=& 2L \sin \ai \, \sin\tau\, \sin\s  \ ,\qquad x_0 =2 L\tau \ ,
\label{hij}
\eea
where $\ai $ is a parameter and $\s\in
[0,2\pi )$.
{}For $\ai $ generic, it describes an ellipse that rotates around
one of its axes and simultaneously performs pulsations, with
one of its radii (the one on the axes of rotation)  becoming zero
at $\tau =n\pi,\ n=$integer.

This string interpolates between the folded string with 
$\ai = 0$, so that
\bea
x_1 &=& 2L \, \cos\tau\, \cos\s  \ ,\qquad x_2=2L\sin\tau\,
\cos\s   \ ,
\nonumber\\
 x_3 &=& 0 \ ,\qquad x_0 =2 L\tau \ ,
\label{hijf}
\eea
and the pulsating circular string with $\ai =\pi/2 $,
\bea
x_1 &=& 0 \ ,  \qquad x_2=2L\sin\tau\,
\cos\s   \ ,
\nonumber\\
 x_3 &=& 2L \, \sin\tau\, \sin\s  \ ,   \qquad x_0 =2 L\tau \ .
\label{hijp}
\eea
At every period of oscillation, there are two times where the string
(\ref{hij}) becomes folded and it can break.
Quantum mechanically, the breaking process is important
during the time that the smaller radius of the ellipse has size $< l_s$.
As it is clear from the above solution, this occurs during
a time $\Delta x_0\cong l_s$ at each period. So the fraction of time
where the string can break is $l_s/L$. This was  used in
 section 1 to  estimate the lifetime by semiclassical arguments.

\subsection{Classical breaking }

Let us now work out the masses of the
outgoing string solutions after the splitting.
The ellipse becomes folded at  $\tau =0$, where
\be
x_1(0, \s)= 2L \cos \ai \,  \cos\s  \ ,\ \qquad x_2(0,\s )=x_3(0,\s )=0\ .
\ee
We assume that the string breaks at $\s=a\pi $ and
$\s=2\pi- a\pi $, $0<a<1$, corresponding to the same spacetime point, because
the string is folded.
Following \cite{IR2}, the outgoing strings I and II and their masses can be determined by
requiring continuity of $x_1, \  x_2,\ x_3 $ and $\dot x_1, \ \dot
x_2,\ \dot x_3 $. 
For the first derivatives, we have
\be
\dot x_1(0,\s )  =0\ , \qquad  \dot x_2(0,\s )  = 2L\cos\s \ , \qquad 
 \dot x_3(0,\s )  = 2L\sin \ai \sin\s \ .
\ee
Now we compute the linear momenta of the outgoing strings I and II.
Since they are conserved, they can be computed at $\tau=0$.
 The components of the momenta are
\be
 p_\I^{x_2}={2\over 2\pi\a' }\int_0^{\pi a} d\s\, \dot  x_2
={4L\over 2\pi\a' }\int_0^{\pi a} d\s\, \cos\s = {2L\over\pi\a' }
\sin\pi a\ ,
\ee
\be
p_\I^{x_1}=0\ ,\qquad p_\I^{x_3}={1\over 2\pi\a' }\Big( \int_0^{\pi a} + \int_{2\pi-\pi a}^{2\pi }\Big)  d\s\, \dot  x_3
=0\ ,
\ee
\be
E_\I= {2\over 2\pi\a' }\int_0^{\pi a} d\s\,  \dot x_0= {2L a\over
\a' }\ .
\ee
Hence
\be
M_\I^2(a)=E_\I^2- (p_\I^{x_2})^2= {4L^2\over {\a'}^2 }\Big( a^2-
{\sin^2\pi a\over \pi^2} \Big)\ .
\label{masuno}
\ee
Similarly
\be
M_{\II }^2(a)= {4L^2\over {\a'}^2 }\Big( (1-a)^2- {\sin^2\pi a\over
\pi^2} 
\Big)\ .
\label{masdue}
\ee
Thus the masses of the two outgoing strings are the same for any $\ai $.
In the case $\ai =0$, we recover the results
of the folded string.
The opposite situation is when $\ai  ={\pi/2}$, 
and we recover the results
of the pulsating circular string \cite{devega}.

In fig. 4 we show the strings I and II at a generic time after the
splitting, for $\theta=\pi/4$ (representing a generic squashing
ellipse). 
The strings I and II rotate and pulsate, and thet have two kinks which
travel along the string. Remarkably, at periodic times, they become
folded.

\bigskip

\begin{figure}[hbt]
\label{fig2} 
\vskip -0.5cm \hskip -1cm
\centerline{\epsfig{figure=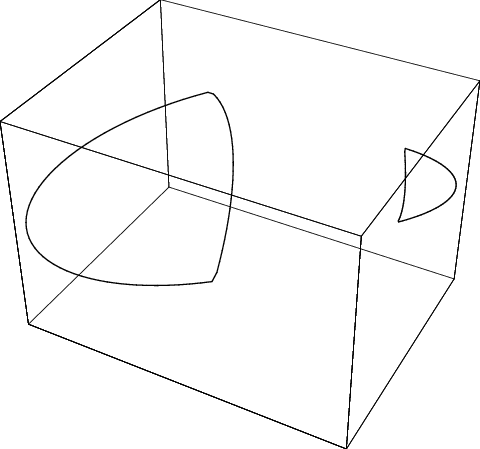,height=6truecm}}
\caption
{\footnotesize Fragments after the splitting of the
squashing string. }
\end{figure}

\subsection{Quantum decay for the squashing ellipse}

The squashing string solution can be
written in terms of Right and Left components as follows
\bea
x_{1L} &=& L \cos \ai \cos(\s_+ )\ ,\qquad x_{1R}= L \cos\ai \cos(\s_- )\ ,
\nonumber\\
x_{2L} &=& L \sin(\s_+ )\ ,\qquad \qquad x_{2R}= L \sin   (\s_- )\ ,
\nonumber\\
x_{3L}&=& -L \sin \ai \cos(\s_+ )\ ,\qquad x_{3R}= L \sin\ai \cos(\s_- )\ ,
\eea
where $\s_\pm=\tau\pm\s $. 
It is convenient to define
\bea
u_R &=& {1\over \sqrt{2}} \big( x_{2R}+i (\cos\ai \, x_{1R}+\sin\ai \, x_{3R})\big)\ ,
\nonumber\\
v_L &=& {1\over \sqrt{2}} \big( x_{2L}+i (\cos\ai \, x_{1L}-\sin\ai \, x_{3L})\big)\ .
\eea

\medskip

In type II superstring theory, one can derive the following
formula for 
$\Delta M^2$ at one loop \cite{IR,CIR}:
\be
\Delta M^2 =c \, g_s^2 \int {d^2 \tau\over \tau_2^3} \int d^2z\ 
\sum_s \langle \hat V^\dagger (z)
\hat V(0)  \rangle _s \ ,
\label{qrates}
\ee
where $\hat V$ is the (super) vertex operator of the string state and
we have indicated the sum over the spin structures $s$.
The decay rate is then obtained as $\Gamma ={\rm Im}\, \Delta
M^2/(2M)$.
[It is also possible --and we have done it in several cases-- to compute the decay rate
 for each of the channels
NS-NS, N-R, R-NS, R-R, that is, for each spin structure separately.]

After evaluating $\langle e^ {-ip\cdot X(z)}  e^ {ip\cdot X(0)}\rangle $,
the world-sheet fermion correlators and summing over
the spin structures, the formula (\ref{qrates}) reduces to
\be
\Delta M^2 =c \, g_s^2 \int {d^2 \tau\over \tau_2^3} \int d^2z\ 
e^{-4 N{\pi ({\rm Im}z)^2\over \tau_2}} \left| {\theta_1(z|\tau )\over
\theta_1'(0|\tau )}\right|^{4N} \langle W^\dagger (z)  W(0)  \rangle \ ,
\label{qrate}
\ee
where, in terms of $u_R,v_L$, the effective bosonic vertex operator
$W$ corresponding to
the quantum state
that describes the squashing ellipse is given by
\be
W= \Big({2\over \a'}\Big)^{N-1} \ N\  V_R V_L\ ,
\ee
\be
V_R= (\p u_R)^{n}\ ,\qquad  V_L=(\bar  \p v_L)^{n}\ ,\qquad n=N-1\ .
\ee

\medskip

Now let us calculate the correlator
\be
\langle   W^\dagger (z)  W(0)  \rangle = 
\langle  (\bar  \p v_L^*(\bar z))^{n} (\p u_R^*(z))^{n} (\p u_R(0))^{n}  (\bar  \p v_L(0))^{n}\rangle\ .
\ee
The non-zero correlators are: 
\bea 
\langle \p u_R^*(z)\p u_R(0)\rangle &=& {\a'\over 2}\big( \p^2\log \theta_1(z) +{\pi\over\tau_ 2}\big)\ ,
\nonumber\\
\langle \bar \p v_L^*(\bar z)\bar \p v_L(0)\rangle &=&
{\a'\over 2}\big(\bar \p^2\log \theta_1(\bar z) +{\pi\over\tau_ 2}\big)\ ,
\nonumber\\
\langle \p u_R^*(z)\bar \p v_L(0)\rangle &=&
\langle \bar \p v_L^*(z)\p u_R(0)\rangle 
={\a'\over 2}\, {\pi\over \tau_2} \cos^2\ai \ ,
\label{ppu}\\
 \langle \bar \p v_L(0) \p u_R(0)\rangle 
&=& \langle \p u_R^*(z)\bar \p v_L^*(\bar z)\rangle 
={\a'\over 2}\, {\pi\over \tau_2} \sin^2\ai \ .
\label{ppy}
\eea
In (\ref{ppy}) we have regularized in a way consistent with (\ref{ppu}).
We obtain
\bea
\langle W^\dagger (z) W(0) \rangle &=& \Big({2\over
\a'}\Big)^{2N\!-\!2}  \sum_{m_1,m_2} { N^2 n!^4\over m_1!^2m_2!^2(n\!-\!m_1\!-\!m_2)!^2}
 \langle \p u_R^*(z)\bar \p v_L^*(\bar z)\rangle ^{m_1}
 \langle \bar \p v_L(0) \p u_R(0)\rangle ^{m_1}
\nonumber\\
&\times & \langle \p u_R^*(z)\bar \p v_L(0)\rangle ^{m_2}
\langle \bar \p v_L^*(z)\p u_R(0)\rangle ^{m_2}
\Big( \langle \p u_R^*(z)\p u_R\rangle 
\langle \bar \p v_L^*(\bar z)\bar \p v_L\rangle \Big)^{n-m_1-m_2}
\nonumber
\eea
Thus
%
\be
\langle W^\dagger (z) W(0) \rangle  = \sum_{l_L,l_R=0}^{N-1}
Q(N,\ai,l_L,l_R)\ \Big( {\pi\over \tau_2}\Big)^{2N-2-l_R-l_L}
(\p^2\log\theta_1(z) )^{l_R} (\bar \p^2\log\theta_1(\bar z) )^{l_L}\ ,
\ee
with
\be
Q(N,\ai,l_L,l_R)=\sum_{m_1,m_2} {N!^2  (\sin\ai)^{4m_1}(\cos\ai)^{4m_2}
\over m_1!^2m_2!^2l_R!l_L! (N\!-\!1\!-\!m_1\!-m_2\!-\!l_R)! 
(N\!-\!1\!-\!m_1\!-m_2\!-\!l_L)! }
\ee
Note that $\theta=0$ and $\theta={\pi\over 2}$ give identical
expressions. This shows the remarkable fact that
the quantum decay rate of the pulsating
circular string and that of the folded string with maximum angular
momentum are identical, confirming the semiclassical
estimate of section 1.

Next, one needs to integrate over $\tau $ and $z$. This is done in two steps 
 \cite{IR,CIR}. One expands the
holomorphic
and antiholomorphic parts of
$\big| {\theta_1(z|\tau )/
\theta_1'(0|\tau )}\big|^{4N} \langle W^\dagger (z)  W(0)  \rangle $
in powers of $e^ {i2\pi\tau }$ and  $e^ {i2\pi z }$. The integrals over
real parts of $\tau $ and $z$ set the same power for each term in the
holomorphic and antiholomorphic parts. The integral over the imaginary
part of $z$ is a gaussian, and the integral over the imaginary part of
$\tau $ is carried out by using 
\be
{\rm Im} \int d\tau_2 \ \tau_2^ {-\alpha } e^ {\tau_2 A}={\pi A^
{\a-1}\over\Gamma(\a )}\ .
\ee
We refer to \cite{IR,CIR} for more details.
The results of the numerical evaluation
of the final expressions are summarized in Table 1,
which shows the contribution of the massive channels
(we have taken $\theta=\pi/4$, as a representative of a typical
squashing string).
Here we are only interested in massive channels, since
we wish to compare with the semiclassical estimate of section 1.

\begin{table}[htbp]
\centering
\begin{tabular}{l|l|l|l|l}
N & $\Gamma _{\rm Jmax}$ & $\Gamma_{\rm  squash}$ \\
 \hline
39 & 4.713 &  0.903  \\
59 & 5.321 & 0.860  \\
79  & 5.718 &  0.822 \\
99 & 6.013 &  0.789  \\
119 & 6.250 &  0.762  \\
\end{tabular}
\parbox{5in}{\caption{Sum over massive channels ($\theta=\pi/4$).
 \label{tab:decaysJ}}} \end{table}

Figure 5 shows the plot of the ratio between $\Gamma_{\rm Jmax}$
and $\Gamma_{\rm squash}$ as a function of the mass $M=\sqrt{N}$. 
According to the semiclassical estimate (\ref{jhg}), (\ref{sq}) of
section 1, 
$\Gamma_{\rm
Jmax}/\Gamma_{\rm squash}$ should depend linearly on $\sqrt{N}$.
We see that this is in perfect agreement with figure 5. Figures 3a and
3b, showing that the lifetime of the squashing string $\Gamma_{\rm
squash}^ {-1} $ is linear
with $\sqrt{N}$,  add more evidence to the law $\Gamma_{\rm squash}\sim N^ {-1/2}$.
Thus the explicit quantum calculation confirms the semiclassical
argument of section 1.

\bigskip

\begin{figure}[hbt]
\label{fig2} 
\vskip -0.5cm \hskip -1cm
\centerline{\epsfig{figure=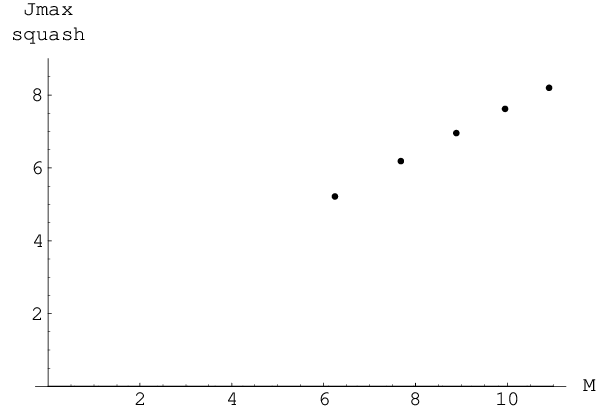,height=6truecm}}
\caption
{\footnotesize
Ratio of decay rates $\Gamma_{\rm Jmax}/\Gamma_{\rm squash}$ 
as a function of $M=\sqrt{N}$. }
\end{figure}

\begin{figure}[hbt]
\label{fig2} 
\vskip -0.5cm \hskip -1cm
\centerline{\epsfig{figure=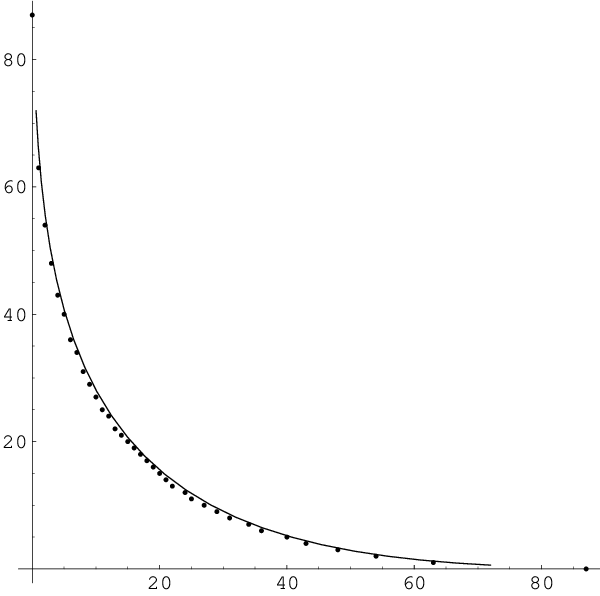,height=6truecm}}
\caption
{\footnotesize Maxima of  $\Gamma(M_\I,M_\II)$
and curve of classical splitting (\ref{masuno}),  (\ref{masdue})
for the squashing string. }
\end{figure}

The quantum decay rate (\ref{qrate}) can be written as a sum over decay
channels $\Gamma(M_\I,M_\II)$ in which the string decays into two fragments of masses
$M_\I$, $M_\II$. We find that, as in \cite{IR}, $\Gamma(M_\I,M_\II)$ is exponentially small
$O(\exp( -cM^ 2))$ except on a curve in the plane $M_\I, M_\II $, where it
has a maximum. This curve precisely corresponds to the curve
(\ref{masuno}),  (\ref{masdue})
determined by the classical splitting, showing that for large $M$ the
classical splitting is the dominant process. 
Figure 6 shows the comparison of the maxima of  $\Gamma(M_\I,M_\II)$
with the classical curve (\ref{masuno}),  (\ref{masdue}) for the case of the squashing string
(the same comparison for the case of Jmax has been done in \cite{IR2}).

\def\k{\kappa }
\def\xe{\zeta }

\section{General class of rotating and pulsating closed strings}
\setcounter{equation}{0}

There is a simple class of solutions of strings rotating in two planes
which contain the cases of the rotating ring, the folded string with maximum angular momentum,
and the squashing ellipses. This class of solutions includes as
particular cases various solutions which have appeared in the
literature (see e.g. recently \cite{turok2}).
In general, the solution represents a string of  elliptical shape
which rotates and pulsates at the same time.
The solution is given by
\bea
Z_1 &=& {X_1+iX_2\over \sqrt{2}}=  L \Big( \cos\ga \cos\gb \ e^{i\s_-}+\sin\ga \sin\gb \
e^{-i\s_-+i\gam  }\nn\\
&+& \sin\ga \cos\gb \  e^{i\s_+}- \cos\ga \sin\gb \
e^{-i\s_+ -i\gam  }\Big) \ ,
\nn\\
Z_2 &=& {X_3+iX_4\over \sqrt{2}}= L \Big( \sin\ga \cos\gb \ e^{i\s_-}-\cos\ga \sin\gb \
e^{-i\s_- +i\gam  } \nn\\
&+&
\cos\ga \cos\gb \  e^{i\s_+}+ \sin\ga \sin\gb \
e^{-i\s_+ -i\gam  }\Big)\ ,
\label{uno}
\eea
$$
X_0=2L\tau= L(\s_++\s_-)\ ,\ \ \ \ \ \s_\pm=\tau\pm\s \ ,\ \ \s\in
[0,2\pi )\ .
$$
The solution is thus characterized by four parameters $(L,\ga ,\ep,
\gam )$.
The mass  of the classical
solution (\ref{uno}) is given by
\be
M={2L\over \alpha '}\ .
\ee
There are six conserved angular momentum components,
\be
J_{ij}= {1\over 2\pi \a' }\int_0^ {2\pi } \big( X_i\dot X_j - X_j\dot X_i)\ .
\ee
We obtain
\be
J_{12}=J_{34}={L^2\over \alpha' } \, \cos 2\gb  \ ,
\ee
\be
J_{24}=J_{31}={L^2\over \alpha' }\, \sin 2\gb\, \sin\gamma \ , 
\ee
\be
J_{14}=J_{23}={L^2\over \alpha' }\, \sin 2\ga\, \ . 
\ee

\medskip

At any given instant of time, the closed string is an ellipse.
By  performing an $SO(4)$ transformation one can go 
to an instantaneous frame where the solution 
takes the form
\be
\tilde Z_1 = L_1(\tau ) \cos (\s-\s_0 ) + i  L_2(\tau )\sin (\s-\s_0 )\ ,\qquad 
\tilde Z_2 =0\ .
\label{das}
\ee
The functions $ L_1(\tau ),\, L_2(\tau )$ give the two radii of the
ellipse at any time. To determine them, we use
\be
Z_1Z_1^*+Z_2Z_2^* =  \tilde Z_1 \tilde Z_1{}^*+ \tilde Z_2 \tilde Z_2{}^* 
= L_1^2(\tau )\cos ^2(\s-\s_0 )+ L_2^2(\tau )\sin ^2(\s-\s_0 )\ ,
\ee
and inserting (\ref{uno}) 
we find
\be
L_{1,2}^2(\tau ) = 2L^2 \big( 1-\cos(\gam ) \cos (2\ga ) \sin(2\gb )\cos(2\tau
) \pm  \sin (2\ga )\sqrt{1 -\cos^2(\gam )\sin^2(2\gb )}\, \big).
\ee
{} The ellipse  contracts to a
minimum size, whose smaller radius is 
\be
L_{2{\rm min}}^2 = 2L^2 \Big( 1-\cos(\gam ) \cos (2\ga ) \sin(2\gb )
- \sin (2\ga )\sqrt{1-\cos^2(\gam )\sin^2(2\gb ) }\, \Big)\ ,
\label{lmin}
\ee
where we assumed $\sin(2\ga  )>0$ and 
$\cos(\gam ) \cos (2\ga ) \sin(2\gb )>0$.

If $L_{2{\rm min}}\neq 0$,
the closed string cannot
break, because there are never two points of the closed string that get
in contact.
In the special case that $L_{2{\rm min}}$ is equal to zero,
 the ellipse contracts to become a folded string, and in this case it
 can break. This is the squashing ellipse discussed in section 3.
If, in addition, $L_{1{\rm min}}=0$, then the string contracts to a
point.
This latter solution  is the pulsating
circular string.

In the case $\gam  =0$ there are some simplifications. One gets
\be
L_{1,2}^2(\tau ) = 2L^2 (1-\cos (2\ga ) \sin(2\gb )\cos(2\tau
)\pm  \sin (2\ga )\cos(2\gb ) )\ .
\ee
We may assume without loss of generality that 
$\sin (2\ga )\cos(2\gb )\geq 0$, since the opposite sign gives
an equivalent solution.
Then $L_2(\tau )\leq L_1(\tau )$. The $\gamma=0 $ ellipse contracts to a minimum
size where the smaller radius is
given by
\be
L_{2{\rm min}}^2=
\begin{cases}
 2 L^ 2 \big( 1-\sin 2(\gb +\ga )\big)  \, , \   &  \cos 2\ga \, \sin 2\gb >0  \ , \cr  
2 L^ 2 \big( 1+\sin 2(\gb + \ga )\big)  \, , \ &  \cos 2\ga \, \sin 2\gb <0 \ .
%
%
\end{cases}
\ee

\medskip
In what follows we discuss different special cases 
separately.

\subsection{Pulsating and rotating rings}

The ellipse becomes a circle, with  $L_1(\tau )=L_2(\tau )$ for all
$\tau $,
in
the cases: a) $\sin 2\ga  =0$ or b) $\gam =0 $ and $\cos(2\gb )=0$.
In this case $L_1(\tau )=L_2(\tau )$.

These  solutions describe a circular string
which, combined with the rotational motion, periodically
contracts to a minimum radius
(which, generically, is different from zero) and then expands to a maximum
radius.

Consider $\ga  =0 $. Then, by introducing new complex coordinates 
$$
\td Z_1 ={Z_1+Z_2^*\over \sqrt{2}}\ ,\qquad \td Z_2 = {Z_1-Z_2^*\over
\sqrt{2}}\ ,
$$
we obtain 
$$
\td Z_1=L_1 \ e^{-i\s }\ \cos(\tau )\ ,\qquad L_1=\sqrt{2} L(\cos\ep 
-e^{-i\gamma }\sin\gb )\ ,
$$
$$
\td Z_2=iL_2\ e^{-i\s }\ \sin(\tau )\ ,\qquad L_2= \sqrt{2}L(\cos\ep
+e^{-i\gamma}\sin\gb )\ .
$$
 It contains the rotating ring  and the pulsating circular
 string 
as the special cases $\beta=0$ and $\beta={\pi\over 4}$,\ $\gamma=0$, respectively.

\subsection{Rotating rigid ellipses}

The ellipses do not pulsate when
\be
\cos(\gam ) \cos (2\ga ) \sin(2\gb )=0\ ,
\label{far}
\ee
which is satisfied for $\gamma={\pi\over 2}$ or $\gb = 0$ or $\ga ={\pi\over 4}$. 
Then $L_1, L_2$ are constants given by
\be
L_{1,2}^2 = 2L^2 \big( 1 \pm \sin (2\ga )\sqrt{1 -\cos^2(\gam )\sin^2(2\gb )}\, \big)\ .
\ee
These rotating rigid ellipses are equivalent to the solutions of \cite{CIR}.
In this work, the quantum decay rate for the corresponding quantum states
was computed by explicit evaluation of ${\rm Im}(\Delta M^2)$ at one loop.



\subsection{Squashing ellipses}

The solution represents an ellipse  that contracts to
a folded closed string, i.e. with 
$L_{2{\rm min}}^2=0$, when
\be
\cos(\gam ) \cos (2\ga ) \sin(2\gb ) + 
\sin (2\ga )\sqrt{1 -\cos^2(\gam )\sin^2(2\gb ) }=1\ .
\ee
The solution of this equation is
\be
\cos(\gam ) ={\cos(2\ga )\over \sin(2\gb )}\ .
\label{cdt}
\ee
The solution discussed in section 3 corresponds to 
$\gam =0$. Then the condition (\ref{cdt}) simplifies to
\be
\sin 2(\gb+ \ga ) =1\ ,  \qquad {\rm or}\ \ \ \ \gb+\ga ={\pi\over 4}\ .
\ee
Introducing new complex coordinates
$$
z_1=x_1+ix_2={Z_1+Z_2\over \sqrt{2}}\ ,\qquad z_2= x_3+ix_4={Z_1-Z_2\over \sqrt{2}}\ ,
$$
and using $\gb={\pi\over 4}-\ga $, we find
\bea
x_1 &=& 2L \sin 2\ga \, \cos\tau\, \cos\s  \ ,\qquad x_2=2L\sin\tau\,
\cos\s   \ ,
\nonumber\\
 x_3 &=& 2L \cos 2\ga \, \sin\tau\, \sin\s  \ ,\qquad x_4\equiv 0
\label{hijz}
\eea
which is equivalent to (\ref{hij}) with $\ai =2\ga+\pi/2$.
 Thus in this special case the string  (\ref{uno}) moves in three spatial dimensions $(x_1,x_2,x_3)$.

\section*{Acknowledgments}

 We would like to thank D. Chialva 
 for  discussions on related topics.
This work is
supported in part by the European
EC-RTN network MRTN-CT-2004-005104. J.R. also acknowledges support by MCYT FPA
2004-04582-C02-01 and CIRIT GC 2005SGR-00564.

\setcounter{section}{0}
\setcounter{subsection}{0}

\setcounter{equation}{0}



\begin{thebibliography}{20}


\bibitem{Dai} J.~Dai and J.~Polchinski,
  ``The Decay Of Macroscopic Fundamental Strings,''
  Phys.\ Lett.\ B {\bf 220}, 387 (1989).


\bibitem{Okada} H.~Okada and A.~Tsuchiya,
  ``The Decay Rate Of The Massive Modes In Type I Superstring,''
  Phys.\ Lett.\ B {\bf 232}, 91 (1989).


\bibitem{CIR} D.~Chialva, R.~Iengo and J.~G.~Russo,
  ``Decay of long-lived massive closed superstring states: Exact results,''
  JHEP {\bf 0312}, 014 (2003)
  [arXiv:hep-th/0310283].

\bibitem{turok}
D.~Mitchell and N.~Turok,
``Statistical Properties Of Cosmic Strings,''
Nucl.\ Phys.\ B {\bf 294}, 1138 (1987).



\bibitem{manes}
J.~L.~Manes,
``Portrait of the string as a random walk,''
JHEP {\bf 0503}, 070 (2005)
[arXiv:hep-th/0412104].


\bibitem{CIR2}  D.~Chialva, R.~Iengo and J.~G.~Russo,
  ``Search for the most stable massive state in superstring theory,''
  JHEP {\bf 0501}, 001 (2005)
  [arXiv:hep-th/0410152].

\bibitem{vilenkin} A. Vilenkin and E.P.S. Shellard, ``Cosmic strings
  and other topological defects'', 
(Cambridge University press 1994). 


\bibitem{DV} T.~Damour and A.~Vilenkin,
``Gravitational wave bursts from cusps and kinks on cosmic strings,''
Phys.\ Rev.\ D {\bf 64}, 064008 (2001)
[arXiv:gr-qc/0104026].


\bibitem{AR} D.~Amati and J.~G.~Russo,
``Fundamental strings as black bodies,''
Phys.\ Lett.\ B {\bf 454}, 207 (1999)
[arXiv:hep-th/9901092].


\bibitem{IR2} R.~Iengo and J.~G.~Russo,
  ``Semiclassical decay of strings with maximum angular momentum,''
  JHEP {\bf 0303}, 030 (2003)
  [arXiv:hep-th/0301109].


\bibitem{DV}
T.~Damour and G.~Veneziano,
  ``Self-gravitating fundamental strings and black holes,''
  Nucl.\ Phys.\ B {\bf 568}, 93 (2000)
  [arXiv:hep-th/9907030].

\bibitem{devega} H.~J.~de Vega, J.~Ramirez Mittelbrunn, M.~Ramon Medrano and N.~Sanchez,
  ``Classical splitting of fundamental strings,''
  Phys.\ Rev.\ D {\bf 52}, 4609 (1995).

\bibitem{IR} R.~Iengo and J.~G.~Russo,
  ``The decay of massive closed superstrings with maximum angular momentum,''
  JHEP {\bf 0211}, 045 (2002)
  [arXiv:hep-th/0210245].


\bibitem{turok2}
G.~Niz and N.~Turok,
``Classical propagation of strings across a big crunch / big bang
singularity,''
arXiv:hep-th/0601007.

\end{thebibliography}
\end{document}